\documentclass[12pt,preprint]{aastex}

\shorttitle{}
\shortauthors{}

\newcommand{\kms}{\rm km\,s$^{-1}$} 
\newcommand{\SFR}{{\rm SFR}}
\newcommand{\myear}{M_\odot\, {\rm yr^{-1}}}
\newcommand{\feii}{\ion{Fe}{2}} 
 
\newcommand{\OIIItwo}{[\ion{O}{3}]\,$\lambda\lambda$4959,5007}

\begin{document}

\title{THE PHYSICAL CONNECTIONS AMONG IR QSOs, PG QSOs AND NARROW-LINE SEYFERT 1 GALAXIES}

\author{
C.N. Hao,\altaffilmark{1} 
X.Y. Xia,\altaffilmark{2,1} 
Shude Mao,\altaffilmark{3,2}
Hong Wu,\altaffilmark{1} 
and Z.G. Deng\altaffilmark{4,1}}

\altaffiltext{1}{National Astronomical Observatories,
                Chinese Academy of Sciences, A20 Datun Road, 100012 Beijing,
                China; Email: hcn@bao.ac.cn.}
\altaffiltext{2}{Dept. of Physics, Tianjin Normal University,
        300074 Tianjin, China.}
\altaffiltext{3}{Univ. of Manchester, Jodrell Bank Observatory,
          Macclesfield, Cheshire SK11 9DL, UK.}
\altaffiltext{4}{College of Physical Science, Graduate School of
        the Chinese Academy of Sciences, 100039 Beijing, China.}   
 
\received{\date}
\accepted{}

\begin{abstract}

We study the properties of infrared-selected QSOs (IR QSOs),
optically-selected QSOs (PG QSOs) and Narrow Line Seyfert 1 galaxies (NLS1s). 
We compare their properties from the infrared to the optical and
examine various correlations among
the black hole mass, accretion rate, star formation rate and
optical and infrared luminosities. We find that
the infrared excess in IR QSOs is mostly in the far infrared, and their
infrared spectral indices suggest that the excess emission is from 
low temperature dust heated by starbursts rather than AGNs. The
infrared excess is therefore a useful criterion to separate the
relative contributions of starbursts and AGNs.
We further find a tight correlation between the star formation rate 
and the accretion rate of central AGNs for IR QSOs. The ratio
of the star formation rate and the accretion rate 
is about several hundred for IR QSOs, but decreases with the central
black hole mass. This shows that the tight correlation between the stellar
mass and the central black hole mass is preserved in massive
starbursts during violent mergers. We suggest that
the higher Eddington ratios of NLS1s and IR QSOs 
imply that they are in the early stage of evolution 
toward classical Seyfert 1's and QSOs, respectively.
\end{abstract}


\section{INTRODUCTION}

  Although much effort has been made since the local ultraluminous
IRAS galaxies (ULIGs) were discovered, the dominant energy output 
mechanism and the evolutionary connection between
circum-nuclear starbursts and active galactic nuclei (AGNs) 
are still a matter of debate
(Kim, Veilleux, \& Sanders 1998; Veilleux, Kim, \& Sanders 1999a; Veilleux,
Sanders, \& Kim 1997, 1999b; Goldader et al. 1995; Murphy et al. 2001; 
Genzel et al. 1998; Lutz et al. 1998; Rigopoulou et al. 1999; Tran et al. 2001).  
 From recent high resolution multi-wavelength observations,
it is now widely accepted that the vast majority of ULIGs 
are strongly interacting or merging galaxies (e.g., Clements et al. 1996;
Murphy et al. 1996; Veilleux, Kim, \& Sanders 2002). The AGN
phenomenon appears at the final merging stage
and the fraction of objects with AGN spectral characteristics
is about 30\% while the fraction of type 1 AGNs is less than 10\% 
(e.g., Clements et al. 1996;  Kim, Veilleux, \& Sanders 1998; Wu et al. 1998; 
Zheng et al. 1999; Canalizo \& Stockton 2001; Cui et al. 2001).
However, the percentage of AGNs increases with 
infrared luminosity, reaching 30--50\% for $L_{\rm IR} > 10^{12.3}L_{\odot}$
(Veilleux, Kim, \& Sanders 1999a). 

Zheng et al. (2002) carefully investigated the optical
spectroscopic properties of infrared-selected type 1 AGNs consisting of 25
objects with infrared luminosities $L_{\rm IR} > 10^{12}L_{\odot}$.
(Following Zheng et al. 2002, we refer to them as IR QSOs.)
They found that the majority of IR QSOs have relatively narrow
permitted emission lines compared with optically-selected Palomar Green quasars
(PG QSOs), which are from the Palomar Bright Quasar Survey Catalogue (Schmidt \& Green 1983) 
with redshift less than 0.5 (Boroson \& Green 1992, hereafter BG92).
Furthermore, more than 70\% of IR QSOs are moderately or extremely strong \feii\
emitters. Canalizo \& Stockton (2001) proposed that such IR QSOs are at
a transitional stage between ULIGs and optically-selected QSOs as their host galaxies 
are undergoing interacting or major merging accompanied
by massive starbursts. In fact, all the optical
spectroscopic properties of IR QSOs show that they
are located at one extreme end of Eigenvector 1 (or the first Principal
Component) defined by BG92 in their principal
component analysis. As pointed out by Grupe (2004), Eigenvector 1
correlates well with the Eddington luminosity ratio $L$/${L_{\rm Edd}}$.
This ratio is thought to be indicative of the `age' of an AGN --
AGNs with a higher Eddington luminosity ratio are at the onset of an
AGN phase. Important clues can therefore be gathered by studying
young forming QSOs with massive starbursts in order to
understand the physics of merging galaxies 
and the AGN phenomenon. Moreover, the co-existence of starbursts and AGNs
provides important information about
the buildup of the stellar populations in
galaxies and the growth of central black holes.

As mentioned above, the fraction of objects with AGN spectral 
characteristics is about 30\% among ULIGs.
High resolution X-ray observations by Chandra and XMM-Newton 
confirmed the existence of central AGNs in some ULIGs through the detection
of ${\rm Fe\, K}\alpha$ lines,  for example in NGC\,6240, Mrk\,273, Mrk\,231 and IRAS\,19254-7545
(Xia et al. 2002; Komossa et al. 2003; Franceschini et al. 2003). 
However, even for AGNs among ULIGs, the dominant energy output
is probably massive starbursts, instead of a central AGN engine.
For example, using the CO kinematic data of IR QSO Mrk\,231, 
Downes \& Solomon (1998) concluded that the central AGN provides only
one third of the total luminosity, with the rest contributed by starbursts.

Massive starbursts may dominate the energy output not only
for some local IR QSOs but also for high redshift massive
starburst galaxies and some optically-selected QSOs. 
A recent deep SCUBA survey uncovered a large population of ULIGs at $z>1$ 
(Wang et al. 2004 and references therein).
Based on ultra-deep X-ray observations and deep optical spectroscopic data,
Alexander et al. (2004) argued that about 40\% of bright SCUBA sources
host AGNs. However, only $\la 20\%$ of the bolometric luminosity
is contributed by AGNs. Recently, Carilli et al. (2004) reported that 
30\% of optically-selected QSOs at high redshift are hyper-luminous
far-infrared galaxies with 
$L_{\rm IR} > 10^{13}L_{\odot}$ and with dust masses $>10^{8}M_{\odot}$. 
These QSOs follow the radio to far-infrared correlation for star-forming
galaxies (Carilli et al. 2001). Therefore, the main energy source for
the SCUBA detected AGNs and some high redshift optical QSOs
may also be starbursts. As the comoving luminosity density of infrared light contributed
by luminous infrared galaxies at $z \sim 1$ is more than 40 times larger than that
in the local universe (Elbaz et al. 2002), it is important to
investigate the properties of these
objects in order to understand the star formation history of the universe
and the number counts of AGNs at high redshift (Alexander et al. 2004).
 Lessons we learn on how to determine the 
dominant energy output mechanism for local IR QSOs
will provide clues to understanding the nature of AGNs
at higher redshifts and the processes involved in
galaxy formation and evolution.

In this paper, we perform statistical analyses for
IR QSOs and compare their properties with
those of PG QSOs and narrow-line Seyfert 1 galaxies (NLS1s).
The outline of the paper is as follows. In \S 2,
we describe how the IR QSO, PG QSO and NLS1 samples are
compiled. In \S 3, we discuss the data reduction and
how we estimate different physical parameters.
The statistical correlations are studied in \S 4. Finally,
in \S 5, we summarize and discuss our results. Throughout
this paper we adopt a cosmology with
a matter density parameter $\Omega_{\rm m}=0.3$, a cosmological constant
$\Omega_{\rm \Lambda}=0.7$ and  
a Hubble constant of $H_{\rm 0}=70\,{\rm km \, s^{-1} Mpc^{-1}}$.

\section{SAMPLE SELECTION}

One aim of our study is to understand the connection of star formation
and accretion process to the central black hole. For this
purpose, we use an infrared-selected type 1 AGN sample as 
the star formation and AGN activity are coeval in these objects.
For comparison, we also compile an optically-selected QSO sample and a NLS1
sample, for which the infrared information is available. 
The details of these three samples are given below:
\begin{enumerate}
\item[(1)]  The infrared-selected type 1 AGN (IR QSO) sample is primarily from 
Zheng et al. (2002). This sample was compiled from the ULIGs in the QDOT redshift
survey (Lawrence et al. 1999), the 1$\,$Jy ULIG survey (Kim \& Sanders 1998), 
and an IR QSO sample obtained by a cross-correlation study of the IRAS Point-Source 
Catalog with the ROSAT All-Sky Survey Catalog. All the IR QSOs selected 
by Zheng et al. are ULIGs with mid-infrared
to far-infrared properties from IRAS observations. 
These galaxies include most of the transition QSOs defined by
Canalizo \& Stockton (2001). Furthermore, they have also been carefully
investigated by Zheng et al. with optical spectra; they concluded
that these objects are in transition from ULIGs to classical QSOs or from 
mergers to elliptical galaxies through a QSO phase.
We added three IR QSOs to the sample obtained from the cross-correlation
of the largest IRAS redshift survey (PSCz)
and the ROSAT archive by Xia et al. (2001). In total,
we have 28 objects, all of which are in the northern sky ($\delta > -30\degr$)
 and they constitute about one third of all the IR QSOs identified using
PSCz, hence it should be a representative sample of IR QSOs.
\item[(2)] The optically-selected QSO sample comprises 57 PG QSOs from
the BG92 sample. For 51 of these,
their infrared information were taken from Haas et al. (2003).
We added six additional objects obtained by a cross-correlation of
87 PG QSOs in BG92 and the IRAS Faint Source Catalog.
\item[(3)] A NLS1 sample was taken from Wang \& Lu (2001) with available IRAS
flux densities from the NED database\footnote{The NASA/IPAC Extragalactic Database (NED) is
operated by the Jet Propulsion Laboratory,
California Institute of Technology,
under contract with the National Aeronautics and Space Administration. }. 
This sample is a heterogeneous sample compiled and observed
spectroscopically by Veron-Cetty et al. (2001). The sample consists of 39
objects, two of which overlap with
the IR QSO sample and eight overlap with the PG QSO sample. In 
addition, the $M_{\rm BH}$ estimation is not available for one of them (MS\,15198$-$0633), so we 
excluded this object from the NLS1 sample, leaving a total of 28 objects.
\end{enumerate}

One complication we already alluded to 
is that there are overlapping objects among the three samples.
For later statistical analyses, we re-group them as follows:
(1) The overlapping objects between the IR QSO and PG QSO samples
are classified in the IR QSO sample. Furthermore, three PG QSOs
(PG\,0050$+$124, PG\,1543$+$489 and PG\,1700$+$518) are re-classified
as IR QSOs following Canalizo \& Stockton (2001) who defined these as in
transition from ULIGs to classical QSOs\footnote{Canalizo \& Stockton
(2001) compiled a sample of transition objects
with nine QSOs and six other similar objects, which were included in the sample
of Zheng et al. (2002).}. 
(2) The original PG QSO sample was selected optically, regardless of the FWHM
of H$\beta$ emission line. In fact, some of the PG QSOs have
FWHM of H$\beta$ less than 2000\kms, satisfying the criteria of
NLS1s\footnote{These PG QSOs were classified as NLS1s in the literature as well.}. 
We will therefore put these objects into the NLS1 sample.
After this exercise, the numbers of objects in the IR QSO, PG 
QSO and NLS1 samples are 31, 41 and 38, respectively.

One criterion of our sample selection is that the objects must
have infrared flux or luminosity information. As the infrared
information is not as readily available as the optical information,
all the three samples used here are somewhat incomplete. However, they
are representative of IR QSOs, optically-selected classical QSOs and NLS1s
in the local universe. Furthermore, as discussed in \S 1, 
there are optical spectroscopic similarities
between IR QSOs and NLS1s and possible evolutionary connections between IR QSOs
and optically-selected QSOs (Sanders et al. 1988a). Therefore,
these three samples, while incomplete, will allow us to explore the physical
relations among these three classes of objects.

\section{DATA REDUCTION AND ESTIMATION OF PHYSICAL PARAMETERS}

In this section, we briefly discuss the data reduction and describe how
we determine the physical parameters of AGNs,
including their black hole masses, infrared and optical luminosities,
H$\beta$ Luminosities,
star formation rates and accretion rates. 
All the parameters are listed in Table 1.

\subsection{Black Hole Masses \label{sec:bh}}
 
The method used to estimate a black hole mass is based on the assumption that
the motion of the gas moving around the black hole is dominated by the
gravitational force and
the broad emission line region (BLR) gas is virialized (see Peterson \& Wandel
1999, 2000 for evidence). Hence the central black hole mass can be estimated
 using the BLR radius and velocity of the BLR gas, i.e.,
\begin{equation}\label{eq:Mbh}
M_{\rm BH} = {R_{\rm BLR} V^2 \over G},
\end{equation}
where ${G}$ is the
gravitational constant. The size of the BLR ($R_{\rm BLR}$) can be estimated from the empirical
relationship between the size and the monochromatic continuum
luminosity at 5100\AA.
This relation was first found by Kaspi et al. (2000) for a sample of 17 Seyfert 
1 galaxies and 17 PG QSOs in a cosmology
with $H_{\rm 0}=75 {\rm km s^{-1} Mpc^{-1}}$, $\Omega_{\rm m}=1$, and
$\Omega_\Lambda=0$.
The relation was refitted in our adopted cosmology by 
McLure \& Jarvis (2002)
\begin{equation} \label{eq:rBLR}
R_{\rm BLR} = (26.4\pm4.4)\left[\frac{\lambda L_{\lambda}(5100{\mbox{\AA}})}{10^{44}\,{\rm erg\,s}^{-1}}\right]^{(0.61\pm0.10)} \mbox{lt-days}.
\end{equation}
The velocity $V$ can be estimated from the Full Width at Half Maximum (FWHM) of the
H$\beta$ broad emission line
$V = \sqrt{3}/2V_{\rm FWHM}$, by assuming that the BLR gas is in
isotropic motions. Therefore, two measurements are needed to
determine the black hole
mass: the luminosity at 5100\AA\ and the FWHM of H$\beta$.
                                                               
We obtained the FWHM of H$\beta$ from Zheng et al. (2002) for 25 out of
the 28 IR QSOs. For the remaining three IR QSOs (IRAS\,F01348$+$3254,
IRAS\,03335$+$4729 and IRAS\,F04505$-$2958),
 we observed them and reduced the
spectra in the same manner as described by Zheng et al. (2002). 
Notice that the FWHM of H$\beta$ was estimated in the same way as in BG92.
The continuum flux densities at 5100\AA\ were measured directly from our
spectra. The uncertainties of the black hole mass were estimated by
error propagation using the uncertainties of the flux density and the FWHM of H$\beta$
measurements given by Zheng et al. (2002). The mean error of the black hole mass
is 0.13dex. This is a lower limit as the uncertainties
of FWHM of H$\beta$ given by Zheng et al. are probably under-estimated (cf. Shemmer et al. 2004)
and there exists other sources of systematic errors (see Wang \& Lu 2001 for more detailed 
discussions). Generally, the black hole mass derived in this way is
accurate within a factor of 2$-$3 (e.g., Wang \& Lu 2001; Marziani et al. 2003, and references
therein; Shemmer et al. 2004).

For the 57 PG QSOs in our sample, the FWHM of H$\beta$ measurements were
from BG92.
The continuum flux densities at 5100\AA\ were taken from the spectrophotometry by
Neugebauer et al. (1987). Specifically, the flux
densities at 5100\AA\ were estimated by a linear interpolation over the
neighboring frequency range.   

For the 28 NLS1s, we used the BLR sizes listed by Wang \& Lu
(2001) to calculate the monochromatic luminosity at 5100\AA\ 
using
the $R_{\rm BLR}-\lambda L_{\rm 5100}$ relation given by Kaspi et al. (2000),
then used eq. (\ref{eq:rBLR}) and 
eq. (\ref{eq:Mbh}) to 
derive the black hole masses from the H$\beta$ FWHM and monochromatic luminosity at 5100\AA.
In this process, we carefully accounted for the difference in the adopted cosmology
(Wang \& Lu 2001 used a cosmology with $H_0=75{\rm km\,s^{-1}}\,{\rm Mpc}^{-1}$,
$\Omega_m=1$ and $\Omega_\Lambda=0$).

\subsection{Infrared Luminosities}
 
For all the sample objects except the 51 PG QSOs, we calculated their infrared
luminosities following Sanders \& Mirabel (1996) based on the flux
densities from the IRAS Faint Source Catalog:
\begin{equation}
L(8-1000\mu {\rm m})=4\pi D_L^2 f_{\rm IR},
\end{equation}
where $D_L$ is the luminosity distance,
and $f_{\rm IR}$ is defined as
\begin{equation}
f_{\rm
IR}=1.8\times10^{-14}\{13.48f_{12}+5.16f_{25}+2.58f_{60}+f_{100}\} {\rm W\,m^{-2}}
\end{equation}
with $f_{12}$, $f_{25}$, $f_{60}$ and $f_{100}$ being the IRAS flux densities at
12, 25, 60 and 100$\mu{\rm m}$ in units of Jy.

Notice that three IR QSOs (IRAS\,06269$-$0543, IRAS\,11598$-$0112 and IRAS\,03335$+$4729)
were not in the IRAS Faint Source Catalog, so we obtained their flux densities from the
IRAS Point Source Catalog. The typical uncertainty of infrared luminosities
is about 0.06 dex. For all the PG QSOs except the six 
objects (i.e., PG\,0923$+$129, PG\,0923$+$201, PG\,1119$+$120, PG\,1351$+$236, PG\,1534$+$580 and PG\,1612$+$261)
obtained by cross-correlating 87 PG QSOs and the IRAS Faint Source Catalog,
their infrared luminosities were calculated by summing over the
$L_{\rm NIR}(3-10\mu {\rm m})$,
$L_{\rm MIR}(10-40\mu {\rm m})$, $L_{\rm FIR}(40-150\mu {\rm m})$ (from ISO observations) and
$L_{\rm sub-mm}(150-1000\mu {\rm m})$ (from the MAMBO and SCUBA
(sub-)millimeter data) in Table 2 of Haas et al. (2003).
There are 16 common objects in the IRAS Faint Source Catalog and Haas et al. (2003).
The average difference of the infrared luminosities measured from the IRAS Faint Source Catalog 
and Table 2 of Haas et al. (2003) for these galaxies 
is 0.014 dex. Hence the infrared luminosities derived in these two different ways
agree well and will not lead to large systematic errors.            

Monochromatic luminosities ($\lambda L_\lambda=\nu L_\nu$) at $12\mu{\rm m}$, $25\mu{\rm m}$, $60\mu{\rm m}$ and
$100\mu{\rm m}$ were also calculated. 
The flux densities in the four bands of the 51 PG QSOs were
derived from Table 1 of Haas et al. (2000, 2003). Notice that some PG
QSOs do not have information in all
four bands\footnote{PG\,1352+183 has no $12\mu{\rm m}$ observation; 
PG\,0007$+$106, PG\,1259$+$593, PG\,1302$-$102, PG\,1307$+$085, PG\,1352$+$183, 
PG\,1411$+$442, PG\,1425$+$267 and PG\,2112$+$059 have no $25\mu{\rm m}$ observations;
PG\,1425+267 has no $60\mu{\rm m}$ observation; PG\,1425$+$267 and PG\,1545$+$210 have no 
$100\mu{\rm m}$ observations.}. For most objects observed by both IRAS and ISO,
the difference of the flux densities at $12\mu{\rm m}$, $25\mu{\rm m}$ and $60\mu{\rm m}$ is within
30\%. However, the difference of flux densities at $100\mu{\rm m}$ is larger 
than 30\% for half of the objects. 
This could be due to the large IRAS beam of about
3$^{'}$ which may enclose cirrus contaminations in 
the flux determinations
(Haas et al. 2003). The difference of flux densities at
$12\mu{\rm m}$, $25\mu{\rm m}$, $60\mu{\rm m}$ and $100\mu{\rm m}$ are 0.072dex, 0.021dex, 0.002dex
and 0.225dex, respectively. However, the larger
errors at $100\mu{\rm m}$ should not statistically 
affect our results that are mainly based on the
flux densities at $60\mu{\rm m}$. Note that no k-correction\footnote{
The k-correction here is defined as the ratio of the intrinsic luminosity
and the observed luminosity.} is applied
to any of the monochromatic or integrated infrared luminosities.
We estimated the k-correction by assuming our objects follow the
spectral energy distribution of either luminous infrared AGNs
or infrared starburst galaxies from Xu et al. (2001). We calculated the k-correction
from $z=0.$ to $z=0.5$ and
found that the largest k-correction occurs for a starburst galaxy 
at $z=0.5$. Even in this case, the k-correction is 
smaller than 0.25dex for the monochromatic luminosity at $25\mu{\rm m}$.  
As all our sample objects have redshifts smaller than 0.5 and many of them 
have spectral energy distributions similar to AGNs, the k-correction
is much smaller than 0.25dex. Hence our results
are not affected significantly by k-corrections.

\subsection{Optical and H$\beta$ Luminosities}
 
The optical luminosities of central AGNs of IR QSOs and PG QSOs were estimated by
the monochromatic continuum luminosity at 5100\AA, 
$L_{\rm opt}=\lambda L_{\lambda}(5100{\mbox{\AA}})$. For PG QSOs, this
approach appears reasonable, as $L_{\rm opt}$'s determined in this way are 
consistent with the $V$-band luminosities obtained through photometric observations.
We will discuss the origin of the optical continuum and the
applicability of this formula to IR QSOs in more detail in \S 4.1.
For IR QSOs, the typical uncertainty of optical luminosities derived in
this way is about 10\%--20\%.

For NLS1s, we used the BLR sizes listed by Wang \& Lu (2001) and 
the $R_{\rm BLR}-\lambda L_{\rm 5100}$ relation given by
Kaspi et al. (2000, cf. eq. \ref{eq:rBLR}) to derive the optical luminosities 
of central AGNs (see the end of \S3.1).

H$\beta$ luminosities were estimated by the H$\beta$ equivalent widths 
and the
monochromatic continuum luminosities at 5100\AA. 
The equivalent widths of the
broad H$\beta$ component for NLS1s were obtained from Veron-Cetty et al. (2001).
For IR QSOs and PG QSOs, the equivalent widths of H$\beta$ were from
Zheng et al. (2002) and BG92, respectively.  Notice that
for IR QSOs and PG QSOs, the
equivalent width of H$\beta$ includes a contribution from a narrow
H$\beta$ component, but as BG92 and Zheng et al. (2002) discussed, this
contribution is less than 3\% of the total H$\beta$ flux,
so the narrow component will not affect the results significantly.
                                                                               
\subsection{Bolometric Luminosities and Eddington Ratios}

For IR QSOs, PG QSOs and NLS1s, we estimated the bolometric luminosities
using (Kaspi et al. 2000),
\begin{equation} \label{eq:Lbol}
L_{\rm bol}\approx9\lambda L_{\lambda}(5100{\rm \mbox{\AA}}).
\end{equation}
Notice that the bolometric luminosity
only refers to the total luminosity associated with the central AGN.

The Eddington ratio $L_{\rm bol}/L_{\rm Edd}$ was then
derived, where the Eddington
luminosity was calculated using the black hole masses 
(e.g., Peterson 1997) already determined in \S\ref{sec:bh}.

\subsection{Accretion Rates and Star Formation Rates}

The accretion rate onto the black hole was calculated by
\begin{equation}
L_{\rm bol}=\eta\dot{M} c^2,
\end{equation}
where $\eta$ is the accretion efficiency and
$\dot{M}$ is mass accretion rate. We adopt 
$\eta=0.1$ throughout this paper (as in many other papers), and we have
\begin{equation}
\dot{M}=6.74\myear {L_{\rm bol} \over 10^{13}L_\odot}.
\label{mdoteq}
\end{equation}
The star formation rate was calculated by the monochromatic 
luminosity at $60\mu {\rm m}$ due to starbursts (see \S 4.3) 
through the following intermediate steps. Notice
that the $60\mu{\rm m}$ flux densities were used for this purpose because 
they have the fewest upper limits.

Lawrence et al. (1989) and Cardiel et al. (2003) gave, respectively,
\begin{equation}
L(40-120\mu{\rm m})\approx 2L_{60\mu{\rm m}}, ~~~
L(8-1000\mu{\rm m}) \approx (1.89\pm0.26)L(40-120\mu{\rm m}).
\end{equation}
So we have
\begin{equation}
L(8-1000\mu{\rm m})\approx 3.78L_{60\mu{\rm m}}.
\label{l60eq}
\end{equation}
Kennicutt (1998) gave an empirical calibration between the star
formation rate and $L(8-1000\mu{\rm m})$:
\begin{equation}
\SFR \approx 4.5 \,\myear {{L}(8-1000\mu{\rm m}) \over 10^{44} {\rm erg\,s^{-1}}}.
\label{sfrcon}
\end{equation}
Combining Eqs. (\ref{l60eq}) and (\ref{sfrcon}), we have
\begin{equation}
\SFR \approx 6.52\, \myear {{L}_{60\mu{\rm m}}\over 10^{10}L_\odot}.
\label{eq:sfreq}
\end{equation}

\section{RESULTS}

Our three samples allow us to investigate the differences of physical properties 
for IR QSOs, optically-selected bright PG QSOs and NLS1s and probe
possible evolutionary connections among these 
objects, and, equally importantly, the interplay between starbursts and AGNs.

We study the correlations between different 
quantities for IR QSOs, PG QSOs and NLS1s. 
For this purpose, we performed
Spearman Rank-order (S-R) correlation analyses. For some objects,
IRAS/ISO observations have only provided upper limits 
of flux densities in one or two bands.  When such data are present
we performed survival analysis\footnote{ASURV, Isobe, Feigelson \& Nelson (1986).}.
The correlation results are listed in Table 2. We discuss
these correlations in more detail below.

\subsection{Correlations\label{sec:correlations}}
  
\begin{figure}
\figurenum{1}
\epsscale{}
\plotone{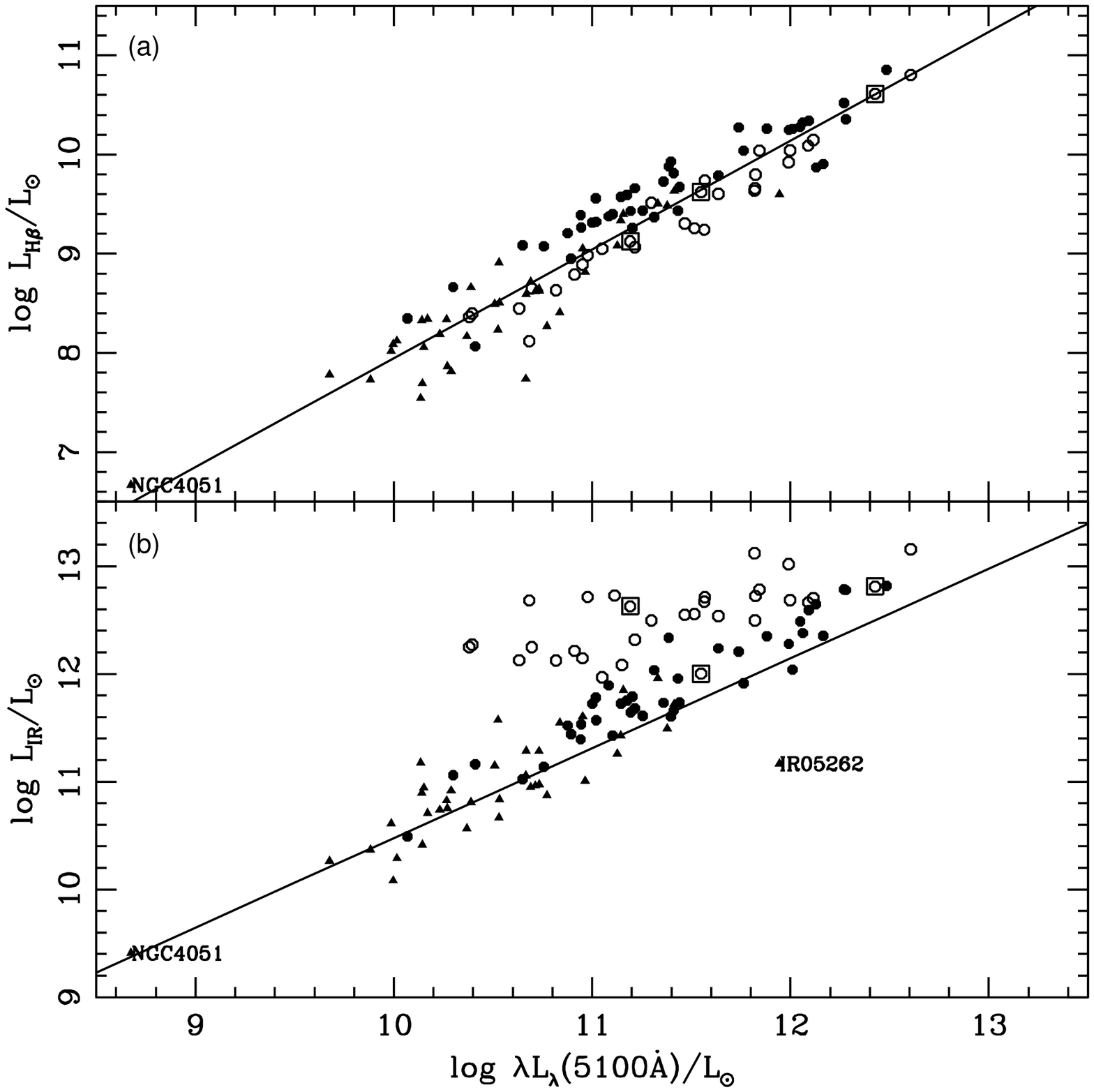}
\caption{(a) H$\beta$ luminosity (b) $L_{\rm IR}$ vs. $\lambda L_\lambda$(5100\AA). 
The open circles represent IR QSOs,
while the filled circles and triangles represent PG QSOs and NLS1s,
respectively. The open circles enclosed by open squares are the three IR
QSOs that are also PG QSOs.
In panel (a), the solid line represents the linear regression for all IR QSOs, PG QSOs and 
NLS1s excluding NGC4051. Note that in all our statistical analysis, NGC4051 is
excluded because it is far from others in terms of all physical parameters. We will not
mention this in later figures.
In panel (b), the solid line represents
the linear regression for all PG QSOs and NLS1s.
For clarity, upper limits of $L_{\rm IR}$ are
not labelled (see Table 1), but survival analyses were performed for
these upper limits.}
\label{lhblirlop.eps}
\end{figure}

Fig.~\ref{lhblirlop.eps}a shows that the featureless optical 
continuum luminosity at 5100\AA\ correlates tightly with the broad H$\beta$ luminosity for
all three samples as a whole. This well-established relation
(Yee 1980; Shuder 1981, see also Osterbrock 1989)
is often used to argue that the predominant mechanism of BLR gas
excitation in AGNs is photo-ionization by the nuclear continuum (e.g., Veilleux,
Kim, \& Sanders, 1999a). Therefore, Fig.~\ref{lhblirlop.eps}a 
suggests that central AGNs 
power the optical emission for IR QSOs, PG QSOs and NLS1s.
This result is consistent with Barthel (2001) who concluded that for QSOs,
the $B$-band magnitude measures the AGN strength. Kauffmann \& Heckman (2004)
also pointed out that the optical continuum of type 1 AGN is dominated by
non-thermal emission. Therefore, it appears reasonable
to adopt a common relation  (eq. \ref{eq:Lbol})
to estimate the bolometric luminosities from the central AGN for all the
objects.

Fig.~\ref{lhblirlop.eps}b 
shows the infrared luminosity versus the optical luminosity 
for PG QSOs, NLS1s and IR QSOs. The solid line shows 
the regression line between these two quantities
for the PG QSOs and NLS1s. One sees that a considerable fraction of IR QSOs
are above the line. If the tight
correlation between the infrared and the optical luminosity for PG QSOs and
NLS1s is because they are both associated with 
central AGNs, then a reasonable 
extrapolation for the infrared excess of IR QSOs is that there is 
another energy source in addition to the AGN that heats the dust.

To further clarify the mechanism that leads to the infrared excess for IR QSOs,
Fig.~\ref{l4band.lop.eps} shows the infrared emission in
four bands (12$\,\mu{\rm m}$, 25$\,\mu{\rm m}$, 60$\,\mu{\rm m}$ and 100$\,\mu{\rm m}$) 
with the optical luminosity. Comparing the four panels,
one can see that the mid-infrared luminosity (at 12$\,\mu{\rm m}$, 25$\,\mu{\rm m}$) 
correlates tightly with the optical luminosity for PG QSOs and
NLS1s, implying that the mid-infrared emission for optically-selected QSOs
and NLS1s is associated with a central AGN. 
The correlations for the far-infrared luminosity (at 60$\,\mu{\rm m}$ and 100$\,\mu{\rm m}$) and optical luminosity 
for PG QSOs and NLS1s are also tight, but
some PG QSOs show deviations (see 
Fig.~\ref{l4band.lop.eps}d). Fig.~\ref{l4band.lop.eps} clearly shows
that IR QSOs deviate from the regression line for PG QSOs and NLS1s
in all four infrared bands; the deviations
are more dramatic in the far-infrared (panels c and d) than
those in the mid-infrared bands (panels a and b).
Therefore, the infrared excess of IR QSOs occurs
mainly in the far-infrared. This is 
further supported by ISO observations which showed
that both PG QSOs and IR QSOs follow similar power-law spectral energy
distributions from the near-infrared 
to mid-infrared band (Haas et al. 2003; Peeters et al. 2004). 
The spectral similarities in the
mid-infrared suggest that the most significant difference
between PG QSOs and IR QSOs must occur in the far-infrared. 

The left panels of Fig.~\ref{his2panel1.eps} show the histogram of $L_{\rm IR}/L_{\rm bol}$
for different samples. As can be seen,
the PG QSOs and NLS1s have similar distributions;
their median values of $L_{\rm IR}/L_{\rm bol}$ are almost the same
($\approx 0.33$), if we assume all the upper limits are real detections.
It implies that roughly one third of the bolometric
luminosity of the optically-selected type 1 AGNs is emitted in the infrared from their dust tori. 
This is consistent with Sanders et al. (1989) who found that all PG QSOs
emit a significant fraction, 10\%--50\% with a typical value of 30\%, of their 
bolometric luminosity in the infrared. In sharp contrast,
more than two thirds of IR QSOs have $L_{\rm IR}/L_{\rm bol}$ ratio larger
than one with a median value of 1.4, highlighting again
the significant infrared excess of the IR QSOs
compared with PG QSOs and NLS1s. 

Given that the infrared excess is mainly in the far-infrared,
we further examine the ratio of the monochromatic luminosity at $60\mu{\rm m}$ to the
bolometric luminosity.
The right panels of Fig.~\ref{his2panel1.eps} show the histograms of this ratio for
IR QSOs, PG QSOs and NLS1s. As can be seen, the median
value (0.50) of $L_{\rm 60 \mu{\rm m}}$/$L_{\rm bol}$ for 
IR QSOs is significantly larger than those (0.09, 0.08) for PG QSOs and
NLS1s, confirming that IR QSOs have high far-infrared excesses compared
with optical QSOs and NLS1s.

\subsection{Eddington Ratios and Spectral Indices}

To investigate the physical connections of IR QSOs, 
optically-selected QSOs and NLS1s, in the following we study
the Eddington accretion ratio, black hole mass and the relation 
of infrared color with infrared excess ($L_{\rm IR} / L_{\rm bol}$).

As is well known, the central engine of an AGN is powered 
by matter accretion onto the black hole. The AGN luminosity is proportional 
to both the mass accretion rate and the accretion efficiency, which is
determined by complex accretion physics. Assuming a fixed
accretion efficiency, the Eddington ratio, $L_{\rm bol} /L_{\rm Edd}$,
measures the accretion rate in units of the critical Eddington value. 

The left panels of Fig.~\ref{his2panel2.eps} show the histograms of $L _{\rm bol}/L_{\rm Edd}$
for three samples. Clearly the IR QSOs have a similar distribution of $L _{\rm bol}/L_{\rm Edd}$ as NLS1s. 
The median values of the Eddington ratio are about 1.7 for IR QSOs and 1.3
for NLS1s, respectively.  More than half of the
IR QSOs and NLS1s have $L_{\rm bol}/L_{\rm Edd}>1$,
implying that the accretion in these systems
may not be spherically symmetric (Collin et al. 2002; Wang 2003).
On the other hand, most PG QSOs have Eddington ratios smaller than one with 
a median value of about 0.2, much smaller than those of IR QSOs and NLS1s. 
As mentioned earlier, IR QSOs and NLS1s have similar optical spectroscopic
properties, and they lie
at one extreme end of the first Principal Component defined by BG92 (Zheng et al. 2002).
This component has been suggested to correlate with $L _{\rm bol}/L_{\rm Edd}$ 
(e.g., BG92; Shemmer \& Netzer 2002).
Therefore the spectroscopic similarities between IR QSOs and NLS1s may
be due to the fact 
that both classes of objects have a high Eddington ratio.

Next we investigate the black hole mass distribution 
(right panels of Fig.~\ref{his2panel2.eps}). It is clear that
NLS1s have much smaller black hole masses compared with IR QSOs and PG QSOs.
The median black hole masses are $5 \times 10^{7}M_\odot$, 
$2 \times 10^{8}M_\odot$ and $6.5 \times 10^{6}M_\odot$
for IR QSOs, PG QSOs and NLS1s, respectively. In addition, the distribution
of black hole mass of IR QSOs is broader than those of PG QSOs and NLS1s, 
which can be explained if 
the black hole masses of IR QSOs are still increasing and have
not yet reached a stable value.

Fig.~\ref{alpha.lirlbol.eps} shows the infrared spectral index of $\alpha (60,25)$
versus the infrared excess of $L_{\rm IR} / L_{\rm bol}$,
where the spectral index is defined as 
\begin{equation}
\alpha(\lambda_1, \lambda_2) =
-{\log(F(\lambda_2)/F(\lambda_1)) \over \log(\lambda_2/\lambda_1)}
\end{equation}
and the wavelengths are in units of $\mu{\rm m}$.
$\alpha(60,25)$ is a measure of the dust temperature (e.g., Sekiguchi 1987). The larger the
value, the higher the dust temperature.
There is a trend from Fig.~\ref{alpha.lirlbol.eps} that as $L_{\rm IR} / L_{\rm bol}$
decreases, $\alpha (60,25)$ increases, implying that the dust temperature increases
as the infrared excess decreases. Statistically, IR QSOs have lower $\alpha
(60,25)$ values and hence
lower dust temperature compared with optically-selected QSOs and
NLS1s. As the dust heated by AGNs tends to have higher temperatures, 
this suggests that starbursts are 
important for heating the dust in IR QSOs.
Our conclusions are supported by preliminary Spitzer 
observations for IR QSO Mrk\,1014 (IRAS\,01572$+$0009, Armus et
al. 2004). These observations
clearly detected the 6.2, 7.7 and 11.3 $\mu{\rm m}$ PAH emission features,
demonstrating convincingly the existence of massive starbursts in IR QSOs.

\subsection{Star Formation Rates and Accretion Rates}

 From our discussions above, IR QSOs are accreting and forming stars at the
same time, below we investigate how these two processes are related to
each other.
Fig.~\ref{l4band.lop.eps}c is a plot of $L_{\rm 60 \mu{\rm m}}$ versus
the accretion rate
$\dot{M}$ for IR QSOs, PG QSOs and NLS1s. 
 From Fig.~\ref{l4band.lop.eps}c, it appears
that $L_{\rm 60 \mu{\rm m}}$ and $\dot{M}$ are correlated for
the IR QSO sample and the combined sample of 
PG QSOs and NLS1s. However, these two sub-samples follow
different regression lines with quite different intercepts
(compare the dashed and solid lines in 
Fig.~\ref{l4band.lop.eps}c).
As we discussed above, the far-infrared excess of IR QSOs are due to
the additional contribution of starbursts to the far-infrared luminosity
compared with PG QSOs and NLS1s.

We can use Fig.~\ref{l4band.lop.eps}c to
estimate the star formation rate of IR QSOs. 
As central AGNs also provide contribution to the
far-infrared emissions for IR QSOs, we first subtract this contribution by
assuming it follows the same
regression relation as PG QSOs and NLS1s (the solid curve in Fig.~\ref{l4band.lop.eps}c).
We then use the excess infrared luminosity at $60\mu{\rm m}$
to calculate the star formation rate using eq. (\ref{eq:sfreq}). We find that
the star formation rate and accretion rate are related to each other by
\begin{equation} \label{eq:sfr}
\log\, \SFR=(0.29\pm0.10)\log\dot{M} +(2.77\pm0.06),
\end{equation}
where $\SFR$ and $\dot{M}$ are both in units of $\myear$. The above
relation can be rewritten as
\begin{equation} \label{eq:sfrmdot}
{\SFR \over \myear}=588.8 \left({\dot{M} \over \myear}\right)^{0.29}.
\end{equation}
In order to examine whether a systematic trend exists between the star
formation rate, accretion rate and black hole mass, in
Fig.~\ref{ir.sbsfrmdot.mbh.eps} we plot 
the ratio of the SFR to $\dot{M}$ versus the black hole mass for IR QSOs.
The regression  line is given by 
\begin{equation} \label{eq:sfrm}
\log {\SFR \over \dot{M}}=(-0.52\pm0.09)\log \left(M_{\rm BH}\over M_{\odot} \right)+(6.62\pm0.71).
\end{equation}
Eqs. (\ref{eq:sfrmdot}) and (\ref{eq:sfrm}) provide important clues about
how the star formation rate and the growth of black holes are connected in
violent merging galaxies. We return to this important 
point in the discussion (\S 5.1).

\section{SUMMARY AND DISCUSSION}

In this paper we have analyzed the statistical properties of IR QSOs, PG QSOs and NLS1s 
from the optical to the infrared. Our results
reveal that these three classes of objects have distinct properties in the infrared. 
Starbursts play a main role in the infrared 
excess, especially the far-infrared excess, in IR QSOs.
Our study also reveals a correlation between the star formation rate
and the accretion rate onto central black holes during galaxy merging and
massive starbursts. This implies that the accretion-driven growth
of central black holes is correlated with the formation of young
stellar population. This has important
implications for the origin of the tight correlation between the
stellar
mass of the hot component of galaxies with the central black hole masses. 
In the following we discuss these issues in more detail.

\subsection{Correlation Between Star Formation Rates and Accretion Rates} 

In the last few years, it has become increasingly clear that the star formation
and AGN activity must be correlated as there are tight
correlations between the black hole mass,
galactic velocity dispersion (e.g. Ferrarese \& Merritt 2000) and
the mass or luminosity of the hot stellar component of the
host galaxy (e.g. Kormendy \& Gebhardt 2001; Magorrian et al. 1998; Laor
1998). It is unclear how the correlations arise. 

By studying a large sample of narrow emission line galaxies
from the Sloan Digital Sky Survey (SDSS), Heckman et al. (2004; see also 
Kauffmann \& Heckman 2004) found that the host galaxies of
bright AGNs have a much younger mean stellar age and quite a large fraction
of these host galaxies have experienced recent starbursts. More importantly,
they found that the volume averaged ratio of the star formation rate to
the  black hole
accretion rate is about 1000 for bulge-dominated galaxies. This value
is in agreement with the ratio of the bulge mass to the
black hole mass empirically
derived (Marconi \& Hunt 2003). Notice that their results are 
based on 23000 narrow emission-line galaxies which excludes type 1
AGNs. In their study, the black hole mass covers more than two orders of magnitude.
Our study, on the other hand, is based on only a few tens of
infrared-selected type 1 AGNs. But these objects are
experiencing both massive starbursts and rapid black hole growth
due to accretions at the same time, and so we probe
the same correlation but in more extreme environments.

At this transitional stage from massive starburst to classical QSO, 
the average ratio of star formation rate to black hole accretion rate 
$\SFR / \dot{M}$ is also a few hundred as shown in Eq.~(\ref{eq:sfrmdot}). 
Comparing Fig.~\ref{ir.sbsfrmdot.mbh.eps} with Figure 11 in Kauffmann \&
Heckman (2004), both the slope and zero-points of
the fitted lines are similar.
It is intriguing that $\SFR/\dot{M}$ is
not constant, but declines with the black hole mass; the same trend
was found in Kauffmann \& Heckman (2004).

On the other hand, as we emphasized in \S4.2, the infrared emission, even
the far-infrared emission, for optically-selected QSOs and NLS1s are not
from the star formation, but mainly from dust heated by central
AGNs. For these objects, the accretion process is still powering
the AGN's emission, but there is no longer substantial
star formation. The picture we obtained here 
is consistent with the
recent simulation result by Springel et al. (2004) that the starburst and AGN activity are coeval, but
the time durations are different as a result of
the detailed form of the response of the gas to the feedback processes.

The number density of ULIGs and classical QSOs are comparable 
in the local universe (Sanders et al. 1988a, 1988b; Canalizo \& Stockton 2001).
The fraction of IR QSOs is less than 10\% of ULIGs, hence 
the number density of IR QSOs in the local universe is at most 10\%
of classical QSOs. If the number density of objects is simply related
to the time scale of different phases, then the time scale for IR QSOs
will be roughly 10\% of that of classical QSOs (about a few times
 $10^{8}$ years, e.g., Marconi et
al. 2004). The IR QSO phase may therefore last only a few times $10^{7}$ years.
The co-moving number density of ULIGs is likely much higher at higher
redshift. For example,
Elbaz et al. (2002) found that the comoving number density of ULIGs is 
several tens times higher at $z \sim 1$ than that in the local universe.
Correspondingly, the co-moving number density of IR QSOs can be higher by the same factor, i.e., 
the co-existing massive starbursts and rapid accretions onto
black holes may be much more common at higher redshift. 
An investigation into the evolution of the co-moving number density of
IR QSOs with redshift will provide valuable information for 
the formation of spheroidals and AGNs.

\subsection{Infrared-Excess as Criterion of Starbursts} 

Based on an analysis for 64 PG QSOs from the infrared to the X-ray,
Haas et al. (2003) concluded that the central AGN is the dominant energy
source for all emissions of PG QSOs. Even for the far-infrared emission,
the central AGN is still the main source of heating on 
the dust tori. If the dust
torus is clumpy, then the central AGN emission
can travel through the gaps farther out and provide the
observed far-infrared emission reradiated
from cooler outer regions. Therefore for PG QSOs, starbursts
play a minor role. In this paper, we showed that NLS1s 
and PG QSOs also satisfy the same correlations and central AGNs
may be the dominant sources from the optical to the far-infrared, 
even for some PG QSOs with $L_{\rm IR} > 10^{12} L_{\odot}$.

The IR QSOs, PG QSOs and NLS1s 
follow the same correlation between
the H$\beta$ luminosity and the optical continuum luminosity 
(see Fig.~\ref{lhblirlop.eps}a and Table 2). This implies that for
all sample objects, regardless of whether they are optically-selected or
infrared-selected, their optical luminosity measures the
central AGN's power. In all the other correlations,
the IR QSOs are significantly different from the PG QSOs and NLS1s,
which is a direct result of the far-infrared excess in IR QSOs.  We
showed that this infrared excess is from starbursts, as can be
most clearly seen from the larger values of  
$L_{\rm IR}/L_{\rm bol}$ and smaller values of $\alpha (60,25)$ (see Fig.~\ref{alpha.lirlbol.eps}).
For most IR QSOs starbursts play a significant,
even dominating, role in their energy output. We conclude that the infrared excess
can serve as an efficient criterion to disentangle the relative
energy contributions of AGNs and starbursts. 

\subsection{Evolutionary Connections Between NLS1s, IR QSOs and PG QSOs}

The optical spectra of IR QSOs and NLS1s are quite similar, both of which have strong or
extremely strong optical \feii\ emission and weak \OIIItwo\ emission.
They are located at one extreme end of the Eigenvector 1 of BG92.
It is widely accepted that the Eigenvector 1 is closely correlated with the
Eddington ratio ($L_{\rm bol} /L_{\rm Edd}$); a large Eddington ratio %
is interpreted as the AGN being at the early stage of an AGN phase,
i.e., having a young `age' (Grupe 2004). As both IR QSOs and NLS1s have high
Eddington ratios (see left panels of Fig.~\ref{his2panel2.eps}), it follows that both
classes of objects are young AGNs in their early stage of evolution.

However, in all the analyses performed in section 4, NLS1s and PG QSOs
have similar
correlations. The differences between them are their
black hole masses and the Eddington ratios. On the other hand, 
IR QSOs and NLS1s are different except that
both have high Eddington ratios. The difference between IR QSOs and
NLS1s is also
underlined by their host galaxies. It appears that the host galaxies of
IR QSOs are merging galaxies while the host galaxies of NLS1s are barred
spirals (Crenshaw et al. 2003). More importantly, IR QSOs are undergoing
massive starbursts which produce the infrared excess in these objects.

\begin{figure}
\figurenum{2}
\epsscale{}
\plotone{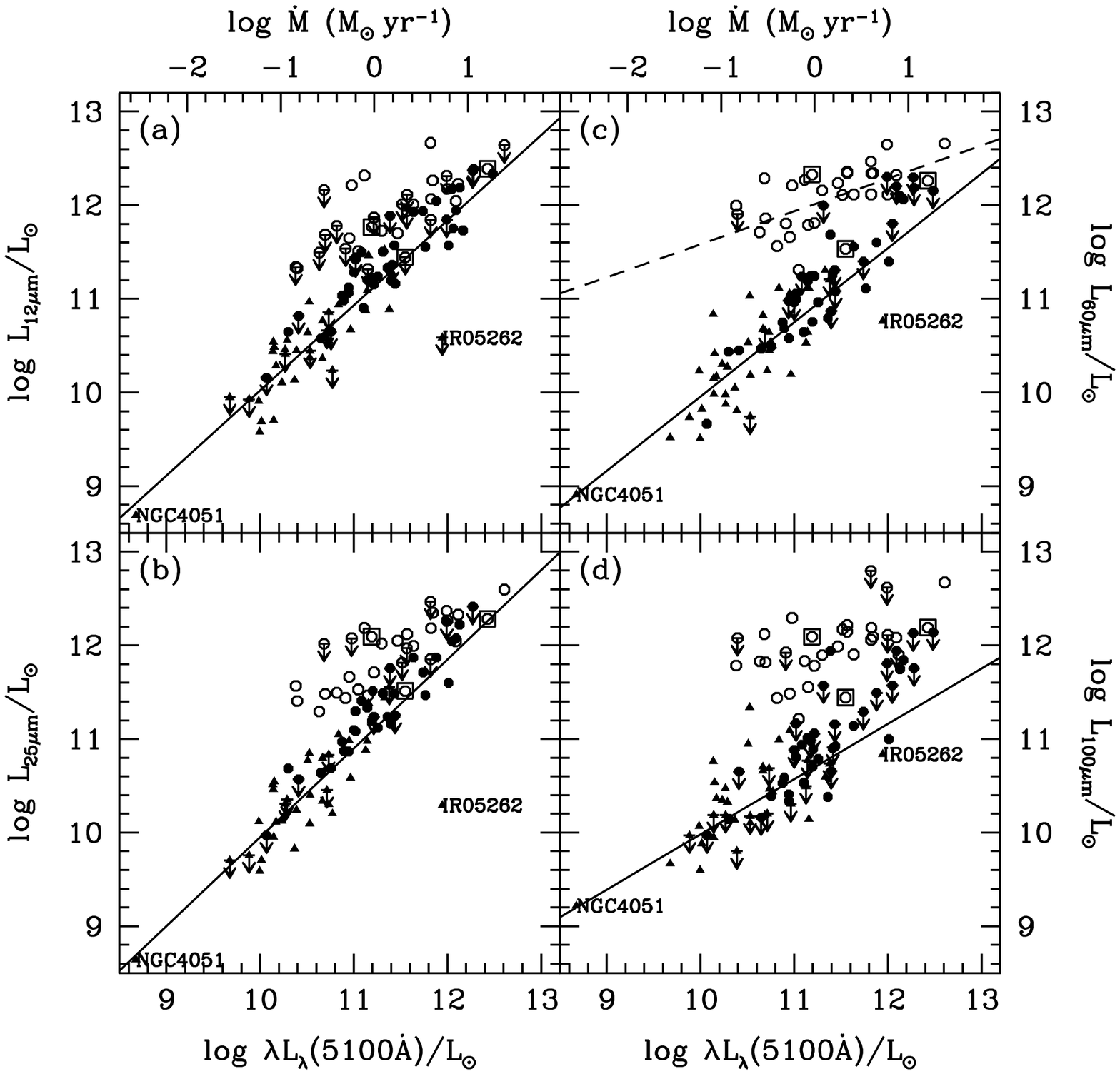}
\caption{The monochromatic luminosities at $12\mu{\rm m}$, 25$\mu{\rm
m}$, $60\mu{\rm m}$ and $100\mu{\rm m}$ vs. $\lambda L_\lambda$(5100\AA). The 
symbols are the same as in Fig.~\ref{lhblirlop.eps} and the arrows denote upper limits. The 
solid line represents the linear regression for all the PG
QSOs with available infrared flux densities and NLS1s. The
accretion rate is labelled at the top abscissa. In panel (c) the best-fit
line for IR QSOs is plotted as the dashed line.}
\label{l4band.lop.eps}
\end{figure}

\begin{figure}
\figurenum{3a}
\epsscale{}
\plotone{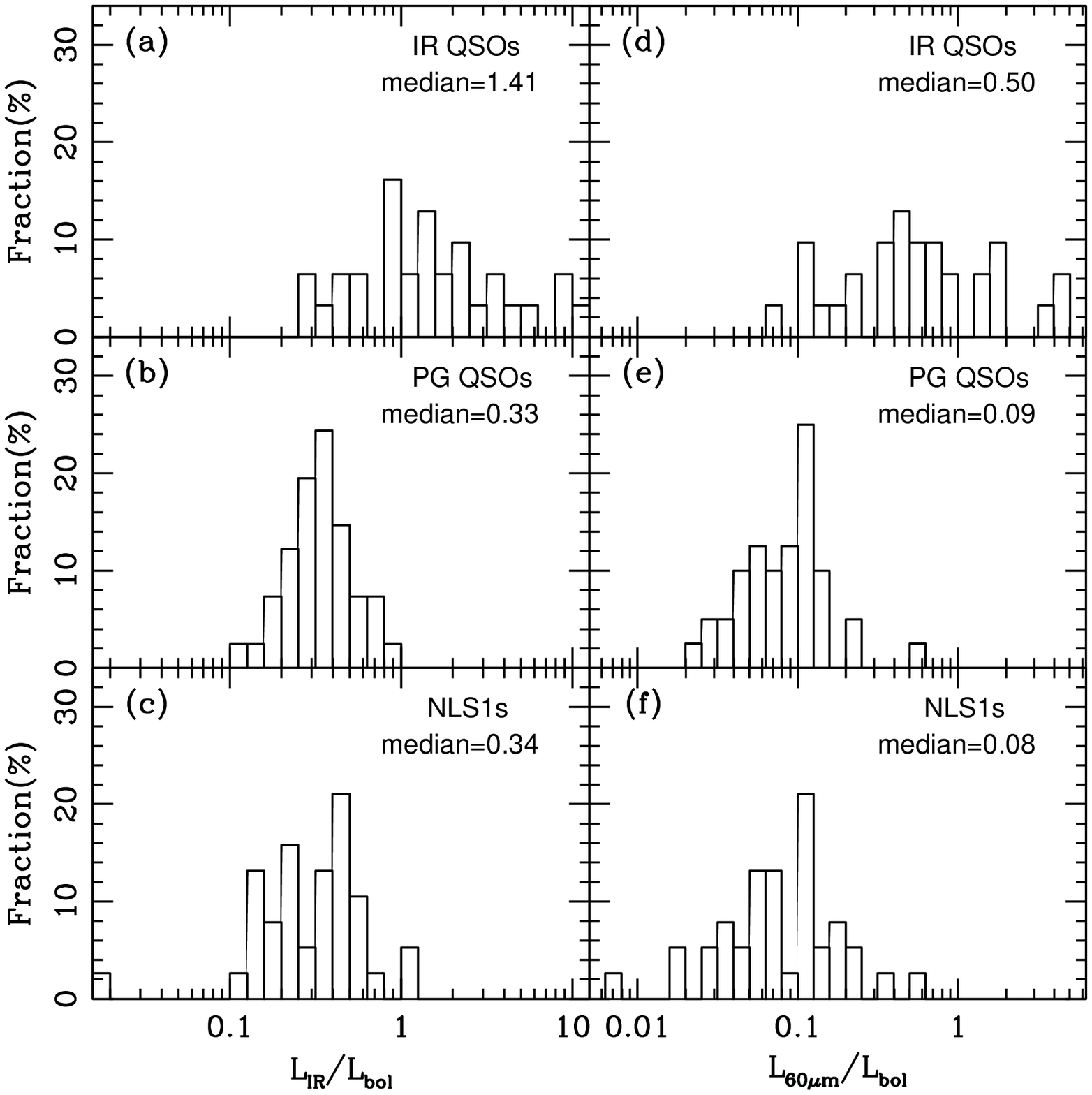}
\caption{Left: Histograms of $L_{\rm IR}/L_{\rm bol}$. 
Right: Histograms of $L_{\rm 60\mu m}/L_{\rm bol}$
for the IR QSOs (top), PG QSOs (middle) and NLS1s (bottom). The median
values are labelled in the panels.
}
\label{his2panel1.eps}
\end{figure}

\begin{figure}
\figurenum{3b}
\epsscale{}
\plotone{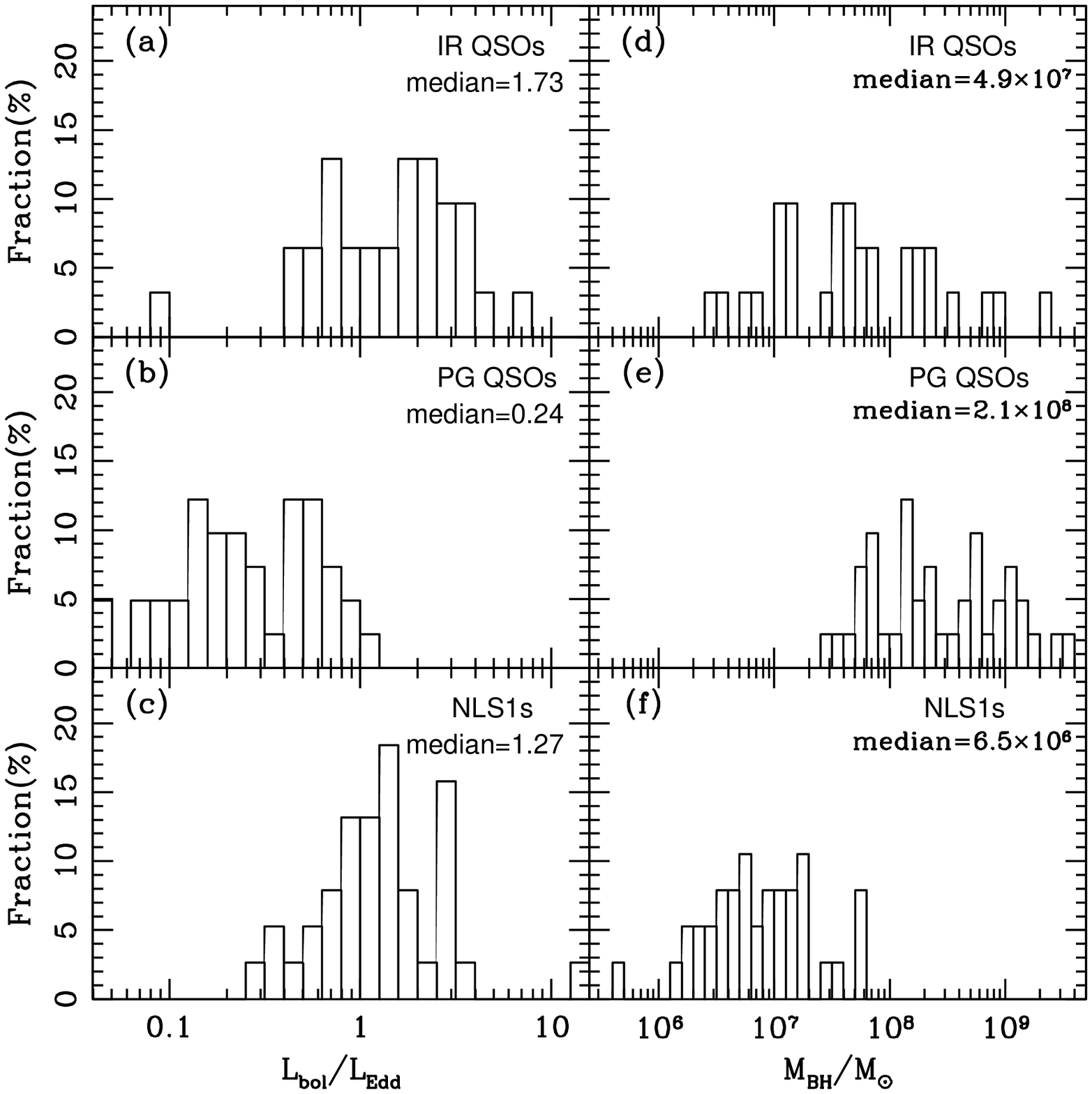}
\caption{Left: Histograms of the Eddington accretion
ratio $L_{\rm bol}/L_{\rm Edd}$.
Right: Histograms of the black hole mass
for the IR QSOs (top), PG QSOs (middle) and NLS1s (bottom). The median
values are labelled in the panels.
}
\label{his2panel2.eps}
\end{figure}

\begin{figure}
\figurenum{4}
\epsscale{}
\plotone{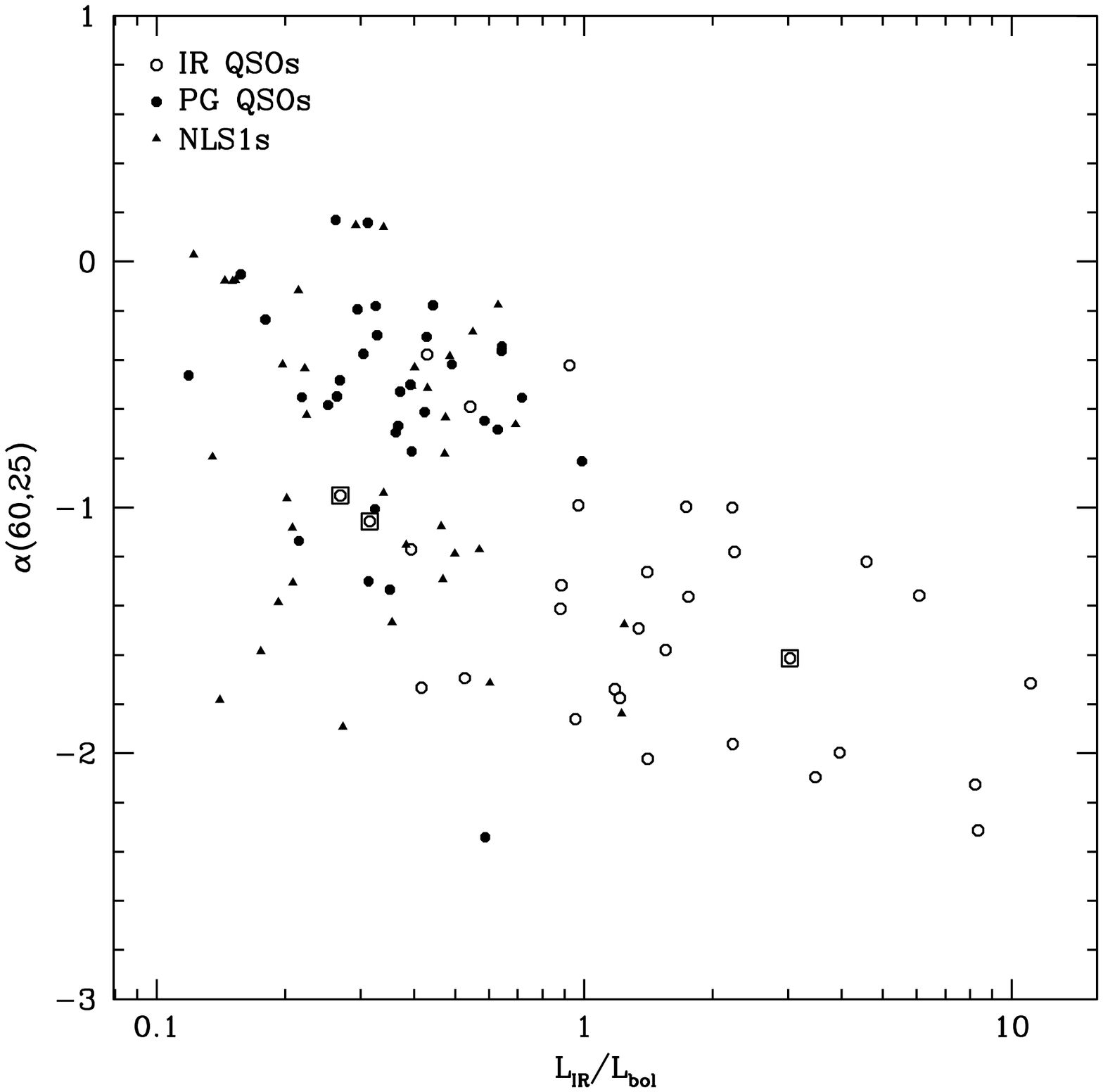}
\caption{The infrared 
spectral index $\alpha (60,25)$ vs. the infrared excess,
$L_{\rm IR}/L_{\rm bol}$.
The symbols are the same as in Fig.~\ref{lhblirlop.eps}.}
\label{alpha.lirlbol.eps}
\end{figure}

\begin{figure}
\figurenum{5}
\epsscale{}
\plotone{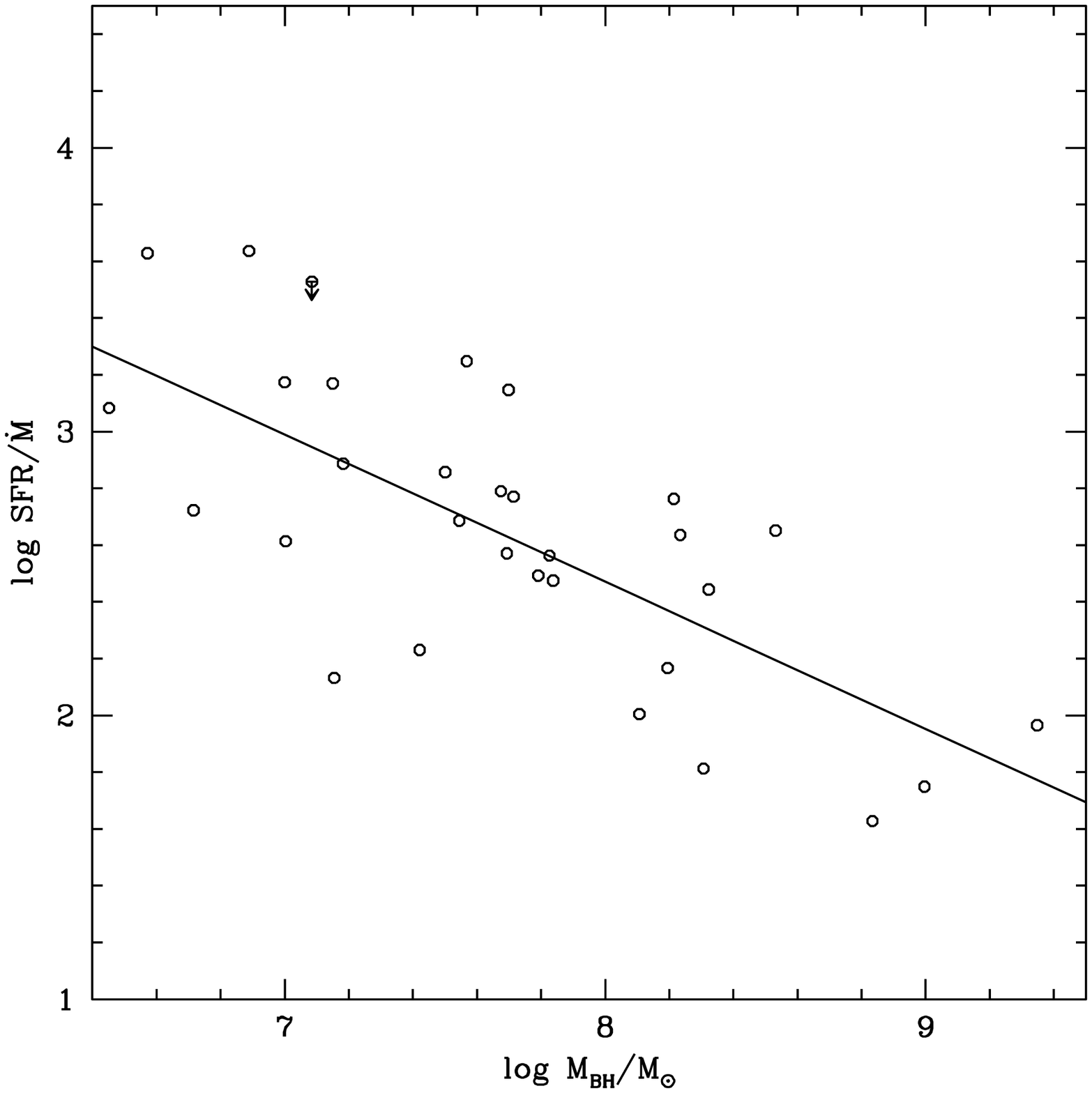}
\caption{The ratio of the star 
formation rate and the accretion rate of black holes vs.
the black hole mass for IR QSOs. The solid line is the best regression
line (see eq. \ref{eq:sfrm}).}
\label{ir.sbsfrmdot.mbh.eps}
\end{figure}

The co-existing massive starbursts and high black hole accretion rate 
in IR QSOs will lead to the rapid growth of
black holes and IR QSOs will rapidly evolve to classical QSOs hosted by elliptical galaxies.
While NLS1s have high Eddington accretion ratios, 
the accretion rates $\dot{M}$ are in fact much smaller than those of IR QSOs 
(see Fig.~\ref{l4band.lop.eps}c), therefore, the final black hole 
for these objects will be smaller, and their
evolution destination may be Seyfert 1s, rather than QSOs.
This is also consistent
with the bulge and black hole mass relation -- 
the bulge mass of spiral galaxies is smaller than those in elliptical
galaxies, and so NLS1s will end as Seyfert 1's with smaller black holes
and hosted by spiral galaxies.

Our analyses provide hints about the physical connections between
different classes of AGNs. It will be important in the future to
quantify how AGNs evolve in the multiple-dimensional space of accretion rate, 
black hole mass, host galaxies etc. With multi-wavelength observations 
and data from large surveys coming in
from both space and ground-based observatories, this task appears to be
increasingly achievable.

\acknowledgments
We would like to thank Drs. X. W. Cao, G. Fazio,
J. S. Huang, J. Y. Wei and X. Z. Zheng for advice and helpful
discussions. Thanks are also due to Dr A. Laor for advice on
how to derive the continuum flux densities for PG QSOs. 
We also thank an anonymous referee and the editor, Dr. J. Shields, for very
constructive comments that improved the paper.
This project is supported by the NSF of 
China No. 10333060, No. 10273012 and TG1999075404.
SM acknowledges the financial support of the Chinese Academy of Sciences
and the hospitalities of Shanghai Astronomical Observatory in particular
Dr. Y. P. Jing during several visits in 2004.

\begin{deluxetable}{ccccccccc}
\tablecolumns{9}
\tabletypesize{\footnotesize}
\tablewidth{0pt}
\tablecaption{Various Physical Parameters}
\tablehead{
\colhead{Name} &
\colhead{Redshift} &
\colhead{log$\left (M_{\rm BH} \over M_\odot \right)$} &
\colhead{log$\left (L_{\rm IR} \over L_\odot \right)$} &
\colhead{log$\left (L_{\rm opt} \over L_\odot \right)$} &
\colhead{log$\left (L_{\rm bol} \over L_\odot \right)$} &
\colhead{log$\left ({L \over L_{Edd}}\right)$} &
\colhead{log$\left (L_{60\mu{\rm m}} \over L_\odot \right)$} &
\colhead{log$\left (L_{\rm H \beta}\over L_\odot \right)$} \\
\colhead{(1)} & \colhead{(2)} & \colhead{(3)} & \colhead{(4)} & 
\colhead{(5)} & \colhead{(6)} & \colhead{(7)} & \colhead{(8)} & \colhead{(9)}}
\startdata
\cutinhead{IR QSOs}
F00275$-$2859 & 0.279 & 7.714 & 12.713\tablenotemark{a} & 11.568 & 12.522 & 0.292 & 12.342 & 9.739 \\
F01572$+$0009 & 0.163 & 7.699 & 12.628 & 11.192 & 12.146 & $-$0.069 & 12.326 & 9.123 \\
F02054$+$0835 & 0.345 & 8.234 & 13.121\tablenotemark{a} & 11.819 & 12.774 & 0.024 & 12.466 & 9.635 \\
F02065$+$4705 & 0.132 & 7.183 & 12.215\tablenotemark{a} & 10.912 & 11.867 & 0.168 & 11.804 & 8.789 \\
F04415$+$1215 & 0.089 & 7.085 & 12.272\tablenotemark{a} & 10.396 & 11.350 & $-$0.250 & 11.905\tablenotemark{b} & 8.395 \\
IR06269$-$0543 & 0.117 & 7.501 & 12.497 & 11.299 & 12.253 & 0.237 & 12.158 & 9.514 \\
F07599$+$6508 & 0.148 & 8.323 & 12.538 & 11.637 & 12.591 & $-$0.248 & 12.116 & 9.603 \\
F09427$+$1929 & 0.284 & 7.568 & 12.715\tablenotemark{a} & 10.977 & 11.931 & $-$0.153 & 12.210 & 8.985 \\
F10026$+$4347 & 0.178 & 7.826 & 12.318\tablenotemark{a} & 11.215 & 12.170 & $-$0.172 & 11.810 & 9.066 \\
F11119$+$3257 & 0.189 & 8.195 & 12.663 & 12.089 & 13.043 & 0.332 & 12.322 & 10.090 \\
Z11598$-$0112 & 0.151 & 6.572 & 12.682\tablenotemark{a} & 10.683 & 11.637 & 0.549 & 12.288 & 8.118 \\
F12134$+$5459 & 0.150 & 6.452 & 12.127\tablenotemark{a} & 10.632 & 11.586 & 0.619 & 11.711 & 8.448 \\
F12265$+$0219 & 0.158 & 8.834 & 12.811 & 12.427 & 13.381 & 0.031 & 12.263 & 10.613 \\
F12540$+$5708 & 0.042 & 8.214 & 12.549 & 11.467 & 12.421 & $-$0.308 & 12.236 & 9.303 \\
F13218$+$0552 & 0.205 & 7.150 & 12.728 & 11.113 & 12.067 & 0.401 & 12.270 & \nodata \\
F13342$+$3932 & 0.179 & 7.421 & 12.496\tablenotemark{a} & 11.821 & 12.775 & 0.838 & 12.116 & 9.657 \\
F15069$+$1808 & 0.171 & 7.000 & 12.249\tablenotemark{a} & 10.696 & 11.651 & 0.135 & 11.861 & 8.651 \\
F15462$-$0450 & 0.101 & 6.889 & 12.250\tablenotemark{a} & 10.381 & 11.335 & -0.070 & 11.995 & 8.361 \\
F16136$+$6550 & 0.129 & 8.996 & 12.003\tablenotemark{a} & 11.550 & 12.504 & $-$1.008 & 11.533 & 9.621 \\
F18216$+$6419 & 0.297 & 9.348 & 13.157\tablenotemark{a} & 12.607 & 13.561 & $-$0.303 & 12.659 & 10.802 \\
F20036$-$1547 & 0.193 & 7.675 & 12.670\tablenotemark{a} & 11.566 & 12.521 & 0.330 & 12.359 & 9.242 \\
F20520$-$2329 & 0.206 & 7.693 & 12.555\tablenotemark{a} & 11.516 & 12.471 & 0.262 & 12.110 & 9.256 \\
F21219$-$1757 & 0.113 & 7.545 & 12.145 & 10.952 & 11.906 & -0.154 & 11.661 & 8.890 \\
F22454$-$1744 & 0.117 & 6.716 & 12.124\tablenotemark{a} & 10.818 & 11.772 & 0.541 & 11.563 & 8.632 \\
F23411$+$0228 & 0.091 & 7.003 & 12.084\tablenotemark{a} & 11.150 & 12.104 & 0.585 & 11.790 & \nodata \\
F01348$+$3254 & 0.367 & 8.532 & 13.018\tablenotemark{a} & 11.991 & 12.946 & $-$0.102 & 12.648 & 9.922 \\
IR03335$+$4729 & 0.184 & 8.107 & 12.686\tablenotemark{a} & 11.998 & 12.952 & 0.330 & 12.112 & 10.043 \\
F04505$-$2958 & 0.286 & 7.791 & 12.723 & 11.824 & 12.778 & 0.471 & 12.341 & 9.798 \\
PG0050$+$124 & 0.061 & 7.155 & 11.970 & 11.050 & 12.004 & 0.333 & 11.310 & 9.050 \\
PG1543$+$489 & 0.400 & 7.838 & 12.784 & 11.843 & 12.797 & 0.443 & 12.344 & 10.039 \\
PG1700$+$518 & 0.292 & 8.307 & 12.703 & 12.115 & 13.070 & 0.247 & 12.090 & 10.148 \\
\cutinhead{PG QSOs}
PG0003$+$158 & 0.450 & 9.068 & 12.784\tablenotemark{a} & 12.270 & 13.224 & -0.359 & 12.298\tablenotemark{b} & 10.521 \\
PG0007$+$106 & 0.089 & 8.416 & 11.429\tablenotemark{a} & 11.103 & 12.058 & -0.874 & 10.645 & 9.400 \\
PG0043$+$039 & 0.384 & 8.992 & 12.279\tablenotemark{a} & 11.992 & 12.946 & -0.561 & 12.303\tablenotemark{b} & 10.248 \\
PG0052$+$251 & 0.155 & 8.639 & 11.735 & 11.441 & 12.395 & -0.759 & 11.080\tablenotemark{b} & 9.673 \\
PG0804$+$761 & 0.100 & 8.131 & 11.733 & 11.360 & 12.314 & -0.333 & 10.794 & 9.728 \\
PG0838$+$770 & 0.131 & 7.841 & 11.571 & 11.020 & 11.974 & -0.383 & 10.994 & 9.321 \\
PG0923$+$201 & 0.190 & 8.936 & 12.335\tablenotemark{a} & 11.386 & 12.340 & -1.112 & 11.684 & 9.880 \\
PG1100$+$772 & 0.313 & 9.133 & 12.040 & 12.011 & 12.965 & -0.684 & 11.396 & 10.258 \\
PG1114$+$445 & 0.144 & 8.309 & 11.893 & 11.084 & 12.038 & -0.787 & 11.236\tablenotemark{b} & 9.376 \\
PG1116$+$215 & 0.177 & 8.319 & 12.206\tablenotemark{a} & 11.739 & 12.693 & -0.142 & 11.398\tablenotemark{b} & 10.274 \\
PG1149$-$110 & 0.049 & 7.481 & 11.061 & 10.299 & 11.253 & -0.744 & 10.435 & 8.663 \\
PG1216$+$069 & 0.334 & 9.008 & 12.487\tablenotemark{a} & 12.049 & 13.003 & -0.521 & 11.804\tablenotemark{b} & 10.281 \\
PG1229$+$204 & 0.064 & 7.841 & 11.139 & 10.756 & 11.710 & -0.647 & 10.492 & 9.074 \\
PG1259$+$593 & 0.472 & 8.778 & 12.778\tablenotemark{a} & 12.279 & 13.233 & -0.061 & 12.187\tablenotemark{b} & 10.357 \\
PG1302$-$102 & 0.286 & 8.712 & 12.353 & 12.165 & 13.120 & -0.108 & 12.063 & 9.905 \\
PG1307$+$085 & 0.155 & 8.640 & 11.660 & 11.411 & 12.365 & -0.791 & 11.257 & 9.811 \\
PG1309$+$355 & 0.184 & 8.138 & 11.957\tablenotemark{a} & 11.433 & 12.387 & -0.267 & 11.305\tablenotemark{b} & 9.433 \\
PG1322$+$659 & 0.168 & 7.985 & 11.611 & 11.256 & 12.210 & -0.291 & 10.962 & 9.435 \\
PG1351$+$236 & 0.055 & 8.209 & 11.163\tablenotemark{a} & 10.411 & 11.365 & -1.360 & 10.450 & 8.065 \\
PG1351$+$640 & 0.087 & 8.567 & 11.789 & 11.203 & 12.157 & -0.926 & 11.248 & 9.259 \\
PG1352$+$183 & 0.158 & 8.158 & 11.754\tablenotemark{a} & 11.176 & 12.130 & -0.543 & 11.244 & 9.592 \\
PG1354$+$213 & 0.300 & 8.362 & 12.035\tablenotemark{a} & 11.312 & 12.266 & -0.612 & 11.997\tablenotemark{b} & 9.368 \\
PG1411$+$442 & 0.089 & 7.757 & 11.533 & 10.945 & 11.899 & -0.374 & 10.579 & 9.263 \\
PG1415$+$451 & 0.114 & 7.709 & 11.442 & 10.894 & 11.848 & -0.377 & 10.683 & 8.950 \\
PG1416$-$129 & 0.129 & 8.752 & 11.606\tablenotemark{a} & 11.397 & 12.351 & -0.917 & 10.868\tablenotemark{b} & 9.930 \\
PG1425$+$267 & 0.366 & 9.533 & 12.379\tablenotemark{a} & 12.062 & 13.017 & -1.032 & \nodata & 10.323 \\
PG1426$+$015 & 0.086 & 8.723 & 11.642 & 11.194 & 12.148 & -1.091 & 10.753 & 9.431 \\
PG1427$+$480 & 0.221 & 7.836 & 11.726 & 11.145 & 12.099 & -0.253 & 11.187 & 9.574 \\
PG1435$-$067 & 0.129 & 8.074 & 11.680 & 11.215 & 12.169 & -0.420 & 11.240 & 9.660 \\
PG1444$+$407 & 0.267 & 8.115 & 12.237 & 11.636 & 12.590 & -0.040 & 11.556 & 9.786 \\
PG1501$+$106 & 0.036 & 8.200 & 11.025 & 10.649 & 11.603 & -1.112 & 10.468 & 9.084 \\
PG1512$+$370 & 0.371 & 9.141 & 12.351 & 11.881 & 12.835 & -0.822 & 11.603 & 10.260 \\
PG1519$+$226 & 0.137 & 7.631 & 11.723\tablenotemark{a} & 11.001 & 11.956 & -0.192 & 10.963\tablenotemark{b} & 9.315 \\
PG1534$+$580 & 0.030 & 7.825 & 10.493\tablenotemark{a} & 10.069 & 11.023 & -1.318 & 9.663 & 8.348 \\
PG1545$+$210 & 0.266 & 9.098 & 11.915 & 11.764 & 12.718 & -0.895 & 11.109 & 10.039 \\
PG1612$+$261 & 0.131 & 7.752 & 11.780\tablenotemark{a} & 11.019 & 11.973 & -0.295 & 11.053 & 9.557 \\
PG1626$+$554 & 0.133 & 8.207 & 11.393 & 10.942 & 11.897 & -0.826 & 10.979\tablenotemark{b} & 9.390 \\
PG1704$+$608 & 0.371 & 9.260 & 12.648 & 12.129 & 13.083 & -0.693 & 12.094 & 9.869 \\
PG2112$+$059 & 0.466 & 8.851 & 12.817\tablenotemark{a} & 12.484 & 13.438 & 0.072 & 12.155\tablenotemark{b} & 10.856 \\
PG2130$+$099 & 0.061 & 7.597 & 11.521 & 10.876 & 11.830 & -0.282 & 10.748 & 9.206 \\
PG2308$+$098 & 0.432 & 9.407 & 12.592\tablenotemark{a} & 12.092 & 13.046 & -0.876 & 12.202\tablenotemark{b} & 10.339 \\
\cutinhead{NLS1s}
TonS180 & 0.062 & 7.091 & 11.259\tablenotemark{a} & 11.127 & 12.081 & 0.474 & 10.530 & 9.082 \\
Mrk359 & 0.017 & 6.329 & 10.416 & 10.144 & 11.098 & 0.253 & 9.982 & 7.692 \\
Mrk1044 & 0.016 & 6.339 & 10.082 & 9.997 & 10.951 & 0.096 & 9.510 & 8.089 \\
IR04312$+$4008 & 0.020 & 6.673 & 10.874\tablenotemark{a} & 10.772 & 11.726 & 0.538 & 10.500 & 8.269 \\
IR04576$+$0912 & 0.037 & 6.581 & 11.178 & 10.136 & 11.090 & -0.007 & 10.835 & 7.542 \\
IR05262$+$4432 & 0.032 & 7.257 & 11.168\tablenotemark{a} & 11.944 & 12.898 & 1.125 & 10.760 & 9.598 \\
Mrk382 & 0.034 & 6.712 & 10.755\tablenotemark{a} & 10.270 & 11.224 & -0.004 & 9.875 & 7.863 \\
Mrk124 & 0.056 & 7.270 & 11.288 & 10.667 & 11.622 & -0.164 & 10.824 & 8.593 \\
Mrk1239 & 0.019 & 6.482 & 10.895\tablenotemark{a} & 10.141 & 11.096 & 0.098 & 10.151 & 8.326 \\
IR09571$+$8435 & 0.092 & 6.801 & 11.573 & 10.525 & 11.479 & 0.163 & 11.029 & 8.232 \\
PG1011-040 & 0.058 & 7.094 & 10.963\tablenotemark{a} & 10.714 & 11.668 & 0.058 & 10.233 & 8.619 \\
PG1016$+$336 & 0.024 & 6.538 & 10.265\tablenotemark{a} & 9.676 & 10.631 & -0.423 & 9.519 & 7.781 \\
Mrk142 & 0.045 & 6.769 & 10.825 & 10.266 & 11.221 & -0.064 & 9.977 & 8.337 \\
KUG1031$+$398 & 0.042 & 6.451 & 10.917\tablenotemark{a} & 10.289 & 11.243 & 0.277 & 10.271 & 7.812 \\
Mrk42 & 0.024 & 6.135 & 10.367\tablenotemark{a} & 9.882 & 10.836 & 0.186 & 9.734 & 7.731 \\
NGC4051 & 0.002 & 5.623 & 9.409 & 8.675 & 9.629 & -0.510 & 8.913 & 6.666 \\
Mrk766 & 0.012 & 6.749 & 10.611 & 9.987 & 10.941 & -0.324 & 10.229 & 8.020 \\
NGC4748 & 0.014 & 6.732 & 10.289 & 10.016 & 10.970 & -0.277 & 9.823 & 8.121 \\
Mrk783 & 0.067 & 7.217 & 11.284\tablenotemark{a} & 10.732 & 11.687 & -0.046 & 10.643 & 8.648 \\
Mrk684 & 0.046 & 6.902 & 10.973 & 10.734 & 11.688 & 0.270 & 10.450 & 8.628 \\
PG1448$+$273 & 0.065 & 6.964 & 11.007\tablenotemark{a} & 10.965 & 11.920 & 0.440 & 10.193 & 8.814 \\
Mrk486 & 0.038 & 7.107 & 10.669\tablenotemark{a} & 10.530 & 11.484 & -0.138 & 9.742\tablenotemark{b} & 8.912 \\
Mrk493 & 0.031 & 6.213 & 10.738 & 10.232 & 11.187 & 0.458 & 10.301 & 8.187 \\
B31702$+$457 & 0.060 & 6.822 & 11.546 & 10.837 & 11.791 & 0.454 & 11.116 & 8.408 \\
Mrk507 & 0.053 & 7.128 & 11.056 & 10.666 & 11.620 & -0.024 & 10.676 & 7.737 \\
HS1817$+$5342 & 0.080 & 7.447 & 11.430\tablenotemark{a} & 11.144 & 12.098 & 0.135 & 10.644 & 9.334 \\
Mrk896 & 0.027 & 6.668 & 10.567 & 10.368 & 11.323 & 0.139 & 10.048 & 8.166 \\
Akn564 & 0.025 & 6.532 & 10.835\tablenotemark{a} & 10.533 & 11.488 & 0.440 & 10.187 & 8.507 \\
PG0003$+$199 & 0.025 & 6.995 & 10.809\tablenotemark{a} & 10.389 & 11.343 & -0.167 & 9.805 & 8.659 \\
PG0026$+$129 & 0.142 & 7.708 & 11.493\tablenotemark{a} & 11.379 & 12.333 & 0.109 & 11.206\tablenotemark{b} & 9.484 \\
PG0923$+$129 & 0.029 & 7.029 & 10.708 & 10.170 & 11.124 & -0.421 & 10.172 & 8.343 \\
PG1001$+$054 & 0.161 & 7.515 & 11.852 & 11.158 & 12.112 & 0.081 & 11.113 & 9.400 \\
PG1119$+$120 & 0.049 & 7.158 & 11.149 & 10.509 & 11.463 & -0.211 & 10.536 & 8.492 \\
PG1211$+$143 & 0.085 & 7.730 & 11.720 & 11.414 & 12.369 & 0.123 & 11.084 & 9.631 \\
PG1244$+$026 & 0.048 & 6.258 & 10.945 & 10.151 & 11.105 & 0.331 & 10.416\tablenotemark{b} & 8.056 \\
PG1402$+$261 & 0.164 & 7.701 & 11.961 & 11.331 & 12.285 & 0.068 & 11.305 & 9.504 \\
PG1404$+$226 & 0.098 & 6.638 & 10.949 & 10.690 & 11.644 & 0.491 & 10.659 & 8.715 \\
PG1440$+$356 & 0.077 & 7.231 & 11.605 & 10.952 & 11.907 & 0.159 & 11.055 & 9.051 \\
\enddata
\tablecomments{Col:(1) name (the prefix IR denotes the IRAS name). Col:(2) redshift. Col:(3) black hole mass. Col:(4) infrared luminosity. Col:(5) monochromatic luminosity at 5100\AA, $\lambda L_\lambda$(5100\AA). Col:(6) bolometric luminosity of AGN (9$\lambda L_\lambda$(5100\AA)). Col:(7) Eddington ratio. Col:(8) monochromatic luminosity
 at 60$\mu$m ($\nu L_\nu$). Col:(9) H$\beta$ luminosity.}
\tablenotetext{a}{The flux density in at least one IRAS band is an
upper limit for objects whose
infrared properties were taken from IRAS; the infrared luminosity
in at least one band is an upper limit for objects from Table 2 of 
Haas et al. (2003).}
\tablenotetext{b}{The $60\mu {\rm m}$ flux density is an upper limit.}
\end{deluxetable}

\clearpage

\begin{deluxetable}{ccccccc}
\tablecolumns{7}
\tabletypesize{\footnotesize}
\tablewidth{0pt}
\rotate
\tablecaption{Significance of Various Correlations}
\tablehead{
\colhead{Relation} &
\colhead{Sample} &
\colhead{Num\tablenotemark{a}} &
\colhead{r$_s$} &
\colhead{Sig(\%)} &
\colhead{a} &
\colhead{b} \\
\colhead{(1)} & \colhead{(2)} & \colhead{(3)} & \colhead{(4)} &
\colhead{(5)} & \colhead{(6)} & \colhead{(7)}}
\startdata
$L_{\rm H\beta}$ vs. $L_{\rm op}$ & IR + PG + NLS1s & 107(0) & 0.945 & $>99.99$ &
1.097$\pm$0.039\phn & -3.026$\pm$0.432\phn \\
~~ & IR QSOs & 29(0) & 0.984 & $>99.99$ &
1.115$\pm$0.051\phn & -3.365$\pm$0.580\phn \\
~~ & PG QSOs + NLS1s & 78(0) &  0.950 & $>99.99$ &
1.140$\pm$0.048\phn & -3.447$\pm$0.531\phn \\
$L_{\rm IR}$ vs. $L_{\rm op}$ & IR QSOs & 31(19) & 0.534 & 99.66 &
0.505$\pm$0.118\phn & 6.547$\pm$1.364\phn \\
~~ & PG QSOs + NLS1s & 78(34) & 0.590 & $>99.99$ &
0.833$\pm$0.063\phn & 2.148$\pm$0.693\phn \\
$L_{\rm 12um}$ vs. $L_{\rm op}$ & IR QSOs & 31(15) & 0.583 & 99.86 &
0.555$\pm$0.126\phn & 5.426$\pm$1.451\phn \\
~~ & PG QSOs + NLS1s & 77(18) & 0.802 & $>99.99$ &
0.910$\pm$0.055\phn & 0.921$\pm$0.606\phn \\
$L_{\rm 25um}$ vs. $L_{\rm op}$ & IR QSOs & 31(6) & 0.737 & 99.99 &
0.510$\pm$0.070\phn & 6.054$\pm$0.794\phn \\
~~ & PG QSOs + NLS1s & 70(14) & 0.794 & $>99.99$ &
0.951$\pm$0.067\phn & 0.440$\pm$0.728\phn \\
$L_{\rm 60um}$ vs. $L_{\rm op}$ & IR QSOs & 31(1) & 0.585 & 99.86 &
0.352$\pm$0.085\phn & 8.062$\pm$0.975\phn \\
~~ & PG QSOs + NLS1s & 77(17) & 0.657 & $>99.99$ &
0.794$\pm$0.062\phn & 2.016$\pm$0.678\phn \\
$L_{\rm 100um}$ vs. $L_{\rm op}$ & IR QSOs & 31(5) & 0.491 & 99.29 &
0.296$\pm$0.090\phn & 8.547$\pm$1.028\phn \\
~~ & PG QSOs + NLS1s & 76(30) & 0.478 & $>99.99$ &
0.591$\pm$0.081\phn & 4.070$\pm$0.884\phn \\
$SFR$ vs. $\dot{M}$ & IR QSOs & 31(1) & 0.491 & 99.29 &
0.291$\pm$0.100\phn & 2.771$\pm$0.059\phn \\
\\
$SFR/\dot{M}$ vs. $M_{\rm bh}$ & IR QSOs & 31(1) & -0.670 & 99.98 &
-0.518$\pm$0.091\phn & 6.615$\pm$0.708\phn \\
\enddata
\tablecomments{Col:(1) relations. Col:(2) sample. Col:(3) number of sources. 
Col:(4) S-R coefficient. Col:(5) significance level. Col: (6) and (7) are the 
coefficients of linear regressions for various correlations obtained using the
EM algorithm (ASURV, Isobe, Feigelson \& Nelson 1986): $\log Y=a\log X+b$.}
\tablenotetext{a}{The number in parenthesis denotes the number of sources whose dependent
variables are upper limits involved in the survival analysis.} 
\end{deluxetable}

\clearpage
 
\clearpage
 
\clearpage
 
\clearpage
 
\clearpage
 
\clearpage

\end{document}